\documentclass[11pt,a4paper]{article}
\pdfoutput=1
\usepackage{jheppub}

\makeatletter
\def\@fpheader{\relax}
\makeatother

\usepackage{url}

\allowdisplaybreaks[2]

\def\cL{\mathcal{L}}

\def\cN{\mathcal{N}}
\def\cO{\mathcal{O}}

\def\cU{\mathcal{U}}

\def\mint{\int_{-\infty}^\infty\!\cdots\!\int_{-\infty}^\infty}

\def\eK{\mathbb{K}}

\def\eP{\mathbb{P}}

\newcommand{\be}{\begin{equation}}
\newcommand{\ee}{\end{equation}}
\newcommand{\ba}{\begin{aligned}}
\newcommand{\ea}{\end{aligned}}

\def\Res{\mathop {\rm Res} \limits}

\DeclareMathOperator{\arccosh}{arccosh}

\def\({\left(}
\def\){\right)}

\def\wt#1{\widetilde{#1}}
\newcommand{\pd}{\partial}

\DeclareMathOperator{\im}{Im}
\DeclareMathOperator{\Tr}{Tr}

\def\mat2#1#2#3#4{\begin{pmatrix} #1 & #2 \\ #3 & #4 \end{pmatrix}}
\def\vec2#1#2{\begin{pmatrix} #1 \\ #2 \end{pmatrix}}

\newcommand{\re}{{\rm e}}
\newcommand{\ri}{{\rm i}}
\newcommand{\rd}{{\rm d}}

\title{Hofstadter's Butterfly in Quantum Geometry}

\author[a]{Yasuyuki Hatsuda,}
\author[b]{Hosho Katsura}
\author[c]{and Yuji Tachikawa}

\affiliation[a]{D\'epartement de Physique Th\'eorique et Section de Math\'ematiques,\\
Universit\'e de Gen\`eve, Gen\`eve CH-1211, Switzerland}
\affiliation[b]{Department of Physics, Graduate School of Science,\\
University of Tokyo, 7-3-1 Hongo, Tokyo 113-0033, Japan}
\affiliation[c]{Kavli Institute for the Physics and Mathematics of the Universe,\\
University of Tokyo, 5-1-5 Kashiwa, Chiba 277-8583, Japan}

\preprint{IPMU16-0078}

\abstract{
We point out that the recent conjectural solution to the spectral problem for the Hamiltonian $H=\re^{x}+\re^{-x}+\re^{p}+\re^{-p}$ in terms of the refined topological invariants of a local Calabi--Yau geometry has an intimate relation with two-dimensional non-interacting electrons moving in a periodic potential under a uniform magnetic field.
In particular, we find that the quantum A-period, determining the relation between the energy eigenvalue and the K\"ahler modulus of the Calabi--Yau, can be found explicitly when the quantum parameter $q=\re^{\ri\hbar}$ is a root of unity,  that its branch cuts are given by Hofstadter's butterfly, and that its imaginary part counts the number of states of the Hofstadter Hamiltonian.
The modular double operation, exchanging $\hbar$ and $\wt{\hbar}=4\pi^2/\hbar$, plays an important role.
}

\begin{document}

\maketitle

\renewcommand{\thefootnote}{\arabic{footnote}}
\setcounter{footnote}{0}
\setcounter{section}{0}

\section{Introduction}

\begin{figure}
\centering
\includegraphics[width=.55\textwidth]{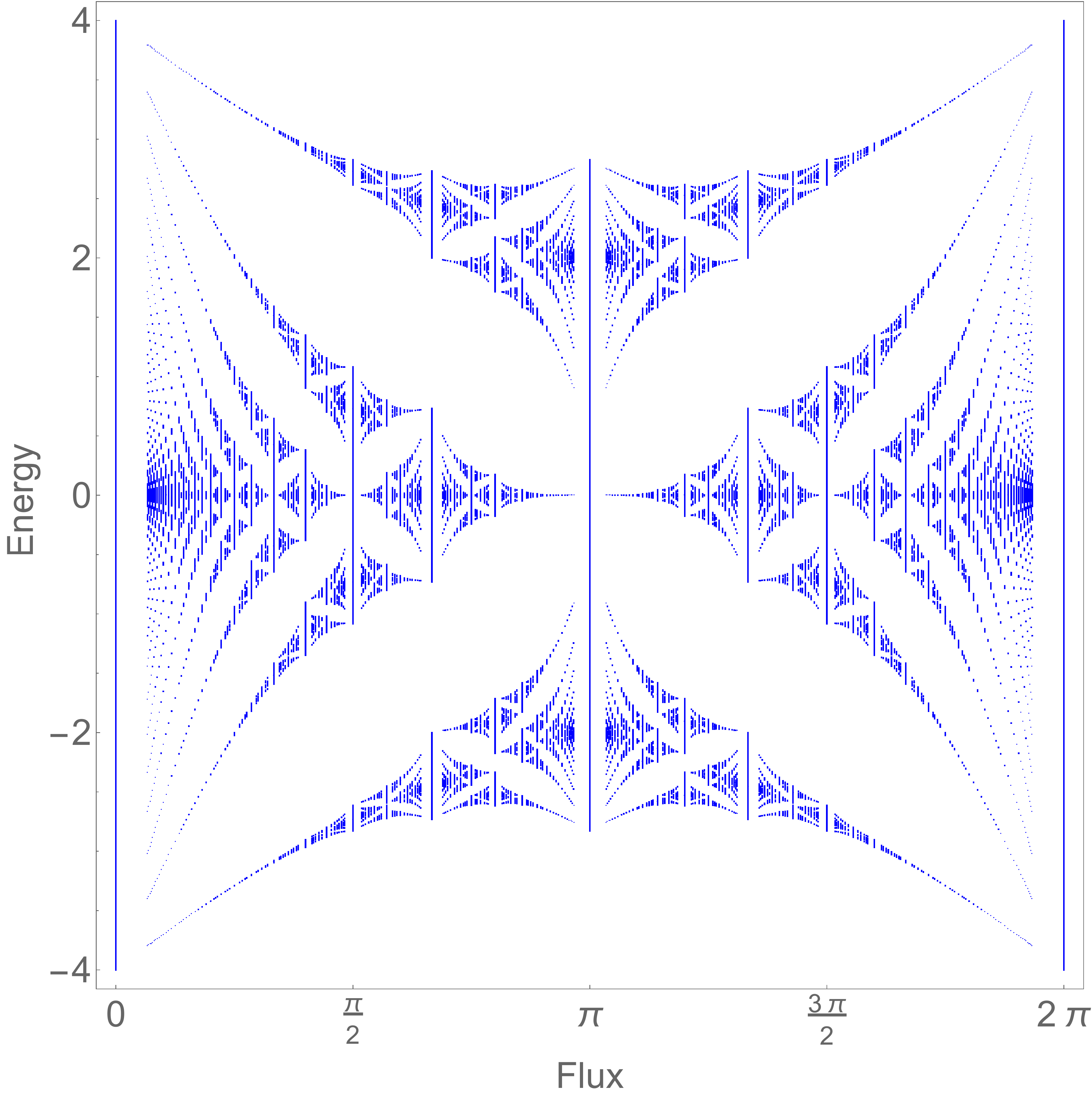}
\caption{The band spectrum of the Hamiltonian \eqref{eq:harper} shows a fractal behavior, called Hofstadter's butterfly.
The vertical direction is the energy, and the horizontal direction is the flux $\hbar$.  
We show the spectrum for $\hbar=2\pi a/b$ with all possible coprime integers $1 \leq a \leq b \leq 30$.
In this paper, we identify this figure with branch cuts of a parameter in a quantum deformed geometry
of a particular Calabi--Yau threefold.  
\label{fig:butterfly}}
\end{figure}

Let us consider the two-dimensional motion of electrons in the presence of the periodic potential and the magnetic field perpendicular to the two-dimensional plane. In suitable limits, the Hamiltonian of the system is described by\footnote{This notation is somewhat unusual
in the context of the two-dimensional (2d) electron system. We will explain a relationship between the parameters here and the standard ones in subsection~\ref{subsec:2d-electron}. We will introduce anisotropy there in the form $H=\re^{\ri x}+\re^{-\ri x}+\lambda(\re^{\ri p}+\re^{-\ri p})$.} \begin{equation}
H=\re^{\ri x}+\re^{-\ri x}+\re^{\ri p}+\re^{-\ri p}, \qquad [x,p]=\ri\hbar.\label{eq:harper}
\end{equation} 
In an old but seminal paper \cite{Hof}, it was found that its spectrum shows an intricate pattern, see Fig.~\ref{fig:butterfly}, now known as Hofstadter's butterfly.
This system was later used  as a model system where the topological numbers determine the Hall conductance \cite{TKNN}.
More recently, this system has received
a renewed interest in the context of ultracold atoms, see e.g.~\cite{PhysRevLett.111.185301, PhysRevLett.111.185302}.
In a more elementary level, one immediately notices that the pattern in Fig.~\ref{fig:butterfly} is self-similar: the Butterfly is a fractal. 
Its combinatorial structure was discussed in detail in e.g.~\cite{PhysRevB.28.6713}.
Note that the spectrum is periodic for $\hbar \mapsto \hbar +2\pi$.
From the figure, we can see that the fractal is apparently generated by transformations \begin{equation}
(\hbar,E) \mapsto (\hbar+2\pi,E),\qquad
(\hbar,E) \mapsto (4\pi^2/\hbar,g(E)), \label{eq:fractal-generator}
\end{equation} where $g(E)$ is an unknown function.\footnote{To the authors' knowledge, neither the explicit form of the function $g(E)$ nor its physical significance is understood in the literature.}

In a completely independent line of research in theoretical high energy physics, the equation\footnote{We can consider a more general equation of the form $H=\re^{x}+\re^{-x}+R^2(\re^{p}+\re^{-p})$, where the parameter $R$ has a natural interpretation in any of the manifestations of this Hamiltonian explained below. We will restore this parameter $R$ in the main part of the paper, while we keep it suppressed during the introduction to reduce the clutter. } \begin{equation}
H=\re^{x}+\re^{-x}+\re^{p}+\re^{-p}, \label{eq:reltoda}
\end{equation} has been intensively studied.
Let us pause here to note that when $x$ and $p$ are restricted to be purely imaginary, this equation reduces to Hofstadter's Hamiltonian \eqref{eq:harper}.

Originally, the variables $x$ and $p$ are regarded as classical complex variables,
and therefore the equation determines a real two-dimensional Riemann surface, or equivalently a complex one-dimensional curve, whose shape is parameterized by the value of $H$.
This  surface arises when mirror symmetry is applied to a non-compact Calabi--Yau geometry known as the local $\mathbb{P}^1\times\mathbb{P}^1$ geometry \cite{Katz:1996fh}, and contains the information on genus-0 Gromov--Witten invariants of the latter. 
Furthermore, the type IIA string theory on this Calabi--Yau geometry is known to give rise to five-dimensional $\mathcal{N}{=}1$ supersymmetric $\mathrm{SU}(2)$ gauge theory compactified on a circle, and as its mirror, the curve knows  the non-perturbative information on this gauge theory \cite{Katz:1996fh}.\footnote{In fact, by replacing $x\mapsto \beta x$ and taking the limit $\beta \to 0$ appropriately,  this curve reduces \cite{Nekrasov:1996cz} to the celebrated Seiberg--Witten curve encoding the information on instantons in $\mathcal{N}{=}2$ supersymmetric pure $\mathrm{SU}(2)$ gauge theory \cite{Seiberg:1994rs}.} 
In \cite{Nekrasov:1996cz}, it was also pointed out that the equation \eqref{eq:reltoda} with the natural Poisson bracket $\{x,p\}_\text{P.B.}=1$ 
is the Hamiltonian of the two-particle relativistic Toda system.

Later, it was appreciated that by elevating $x$ and $p$ in \eqref{eq:reltoda} to quantum variables satisfying the commutation relation 
$[x,p]=\ri \hbar$, we can extract more information both on the said $\mathrm{SU}(2)$ gauge theory \cite{Nekrasov:2009rc, Nekrasov:2013xda} and on the topological invariants of the local Calabi--Yau geometry \cite{Aganagic:2011mi, Huang:2014nwa}. 
In the last few years, it was finally realized that we can conversely use the topological invariants of the Calabi--Yau geometry, which can be computed independently, to describe the eigenvalues of the Hamiltonian \eqref{eq:reltoda}, where $x$ and $p$ are taken to be real.
The complete eigenvalues are determined by an exact version of the Bohr--Sommerfeld quantization condition \cite{Grassi:2014zfa},
based on earlier attempts \cite{Kallen:2013qla, Huang:2014eha}. 
This quantization condition has not yet been rigorously proven, but passes extensive analytical and numerical tests \cite{Kashaev:2015kha, Marino:2015ixa, Kashaev:2015wia, Wang:2015wdy, Gu:2015pda, Codesido:2015dia, Hatsuda:2015fxa, Hatsuda:2015qzx, Franco:2015rnr, Okuyama:2015pzt, Bonelli:2016idi, Kashani-Poor:2016edc, Grassi:2016vkw}.\footnote{The quantization condition was originally given as the formula for the spectral determinants \cite{Grassi:2014zfa}. The history up to this point is nicely summarized in the review paper \cite{Marino:2015nla}. A conjecturally equivalent but distinct form was proposed in \cite{Wang:2015wdy}, whose invariance under $\hbar\leftrightarrow \wt{\hbar}=4\pi^2/\hbar$ was emphasized in \cite{Hatsuda:2015fxa}, and applied to relativistic Toda systems in \cite{Hatsuda:2015qzx} 
and further to a wide class of integrable systems \cite{Franco:2015rnr}. In this paper we utilize this latter form.}
We should note that there is a parallel development purely in the 5d gauge theoretic framework \cite{Gaiotto:2014ina, Bullimore:2014upa, Bullimore:2014awa}.

To write down the quantization condition, we first need a function of the energy $E$ known as the quantum A-period of the geometry:\begin{equation}
t=t(E,q), \qquad  (q=\re^{\ri\hbar}), \label{eq:A-period}
\end{equation} which is explicitly computable \cite{Fucito:2011pn, Aganagic:2011mi}.
Then, the $n$-th energy eigenvalue $E_n$ is given by the  exact Bohr--Sommerfeld condition \begin{equation}
\frac{\partial}{\partial t} F(t,q)
+\frac{\partial}{\partial \wt{t}}\, F(\wt{t}, \wt{q})
=2\pi\left(n+\frac12\right),\label{eq:ExactQC}
\end{equation}
where the tilded variables are defined by \begin{equation}
\wt{q}=\re^{\ri\wt{\hbar}}, \qquad
\wt{\hbar} = \frac{4\pi^2}{\hbar}, \qquad \wt{t}=\frac{2\pi t}{\hbar}, 
\end{equation} and 
$F(t,q)$ is another function explicitly computable from the Calabi--Yau geometry, essentially given by the free energy of the topological string on this geometry in the Nekrasov--Shatashvili limit \cite{Nekrasov:2009rc}.

We note that the quantum A-period \eqref{eq:A-period} is invariant under the transformation
\be
T:\hbar \mapsto \hbar+2\pi.\label{eq:T}
\ee
Surprisingly the quantization condition \eqref{eq:ExactQC} is invariant under another transformation \begin{equation}
S: \hbar \mapsto \wt{\hbar} =\frac{4\pi^2}{\hbar}.\label{eq:S}
\end{equation}
Note also that these two transformations are just related to the fractal-generating transformations \eqref{eq:fractal-generator}.

The aim of this paper is twofold: one is to extract out more on the relativistic Toda spectrum from this invariance under $S: \hbar\leftrightarrow \wt{\hbar}$, and another is to explore its possible relation with Hofstadter's butterfly. 
Our strategy is to restrict $\hbar$ to be a rational multiple of $\pi$, or equivalently to take $q$ to be a root of unity. 
In Sec.~\ref{sec:double}, we use the idea of the modular double to show that the energy eigenvalues $E_n$ and $\wt{E}_n$ of the relativistic Toda system, whose Planck constant is respectively $\hbar$ and $\wt{\hbar}$, satisfy an algebraic relation of the form $P_{a/b}(E_n)=P_{b/a}(\wt{E}_n)$ 
for $\hbar=2\pi a/b$, where $P_{a/b}(x)$ is a degree-$b$ polynomial.
In Sec.~\ref{sec:hof}, we first see that the transformations \eqref{eq:T} and \eqref{eq:S} 
allow us to determine the quantum A-period $t(E,q)$ in a closed form expression if $q$ is a root of unity,
where the most of the $q$ dependence is encoded in the polynomial $P_{a/b}(E)$.
We will then note that the same polynomial $P_{a/b}(E)$ determines Hofstadter's butterfly by the equation $-4 < P_{a/b}(E)< 4$. 
In particular, the branch cuts of $t(E,q)$ are exactly on the energy bands of Hofstadter's butterfly.
In addition, we will show that the imaginary part of the quantum A-period is precisely the integrated density of states of the Hofstadter Hamiltonian.
We will conclude the paper with a short discussion in Sec.~\ref{sec:conclusions}.

\section{Modular double in the relativistic Toda lattice}\label{sec:double}
Let us start with a hidden duality existing in the relativistic Toda lattice.
This duality is called the modular double, first found in quantum groups \cite{Faddeev:1995nb} (see also \cite{Faddeev:1999fe}).
The modular double also appears, for instance, in 2d CFTs, in gauge theories and in integrable systems.
It is argued in \cite{Kharchev:2001rs} that the relativistic Toda lattice has the modular double associated with
$\cU_q(sl(2,\mathbb{R}))$.
We will show that the modular double directly relates the spectrum at the Planck constant $\hbar$
to that at $\wt{\hbar}=4\pi^2/\hbar$ if $q=\re^{\ri \hbar}$ is a root of unity.%
\footnote{Very recently, the modular double property is also argued 
to explore exact eigenfunctions in the relativistic Toda lattice \cite{Sciarappa:2016ctj}. We also note that an explicit construction
of eigenfunctions is presented in \cite{Marino:2016rsq}.}

Although we can present the analysis for general number $N$ of the particles, we here keep $N=2$ for simplicity. The general case can be treated similarly.
The Hamiltonian of the periodic relativistic Toda lattice with just $N=2$ particle, after removing the center-of-mass mode, is given by 
\be
H=R^2(\re^{p}+\re^{-p})+\re^{x}+\re^{-x},\qquad [x,p]=\ri \hbar.
\label{eq:H-N=2}
\ee 
The basic concept of the modular double is that there is a \textit{dual} Hamiltonian, which
is obtained from $H$ by exchanging $\hbar \leftrightarrow \wt{\hbar}=4\pi^2/\hbar$:
\be
\wt{H}=\wt{R}^2(\re^{\wt{p}}+\re^{-\wt{p}})+\re^{\wt{x}}+\re^{-\wt{x}},\qquad [\wt{x},\wt{p}]=\ri\wt{\hbar},
\label{eq:H-dual-N=2}
\ee
where
\be
\wt{p}=\frac{2\pi p}{\hbar},\qquad \wt{x}=\frac{2\pi x}{\hbar}, \qquad 
\wt{R}=R^{2\pi/\hbar}.
\ee
The important point is that the original Hamiltonian and its dual commute:
\be
[H,\wt{H}]=0.
\ee
Thus one can diagonalize these two simultaneously:
\begin{align}
H Q(x) = E Q(x), \qquad \qquad 
\wt{H} Q(x)= \wt{E} Q(x).
\label{eq:eigen-system}
\end{align}
These eigenvalues take discrete values, when the wavefunction is considered as a square-integrable function on the real line.
At first glance, it is far from obvious how these two eigenvalues $E$ and $\wt{E}$ are interrelated.
In \cite{Hatsuda:2015qzx}, exact quantization conditions that determine all the eigenvalues of the relativistic Toda lattice of $N$-particle were conjectured.
One important consequence in \cite{Hatsuda:2015qzx} was that these quantization conditions are invariant under the
S-transform \eqref{eq:S}, which implicitly relates the eigenvalues $E$ to the dual ones $\wt{E}$,
in terms of the quantum A-period.
We will see just below that the modular double relates these two eigenvalues more directly
if $q$ is a root of unity.

In this case of $N=2$, we see that the eigenvalue equations \eqref{eq:eigen-system} immediately give the difference equations
\be
\ba
Q(x+\ri \hbar)+Q(x-\ri \hbar)
&=T(x) Q(x), \\
Q(x+2\pi \ri)+Q(x-2\pi \ri)
&=\wt{T}(x) Q(x),
\ea
\label{eq:Baxter-N=2}
\ee
where
\be
T(x)=R^{-2}\(E-2\cosh x  \),\qquad
\wt{T}(x)=\wt{R}^{-2}\(\wt{E}-2\cosh \frac{2\pi x}{\hbar} \).
\ee
Note that the eigenfunction $Q(x)$ in these difference equations is the \textit{same} function.
This fact is crucially important in our analysis below.
Note that the T-functions have the following periodicity:
\be
T(x+2\pi \ri)=T(x),\qquad \wt{T}(x+\ri \hbar)=\wt{T}(x).
\ee
Here we pause to mention that even in the general case of more particles $N>2$, the relativistic Toda lattice can be reduced to a similar one-dimensional problem via  Sklyanin's separation of variable method, see e.g.~\cite{Kharchev:2001rs}. In this context the equations \eqref{eq:Baxter-N=2} are called the Baxter and dual Baxter equations, respectively.

Now we show that if the Planck constant takes the form
\be
\hbar=2\pi \tau ,\qquad \tau=\frac{a}{b},
\label{eq:hbar-cond}
\ee
with coprime integers $a$ and $b$, then the two Baxter equations \eqref{eq:Baxter-N=2}
lead to a non-trivial relation between $E$ and $\wt{E}$.
The condition \eqref{eq:hbar-cond} is rephrased as saying that the quantum parameter $q=\re^{\ri \hbar}$ is a root of unity.
Shifting $x \to x+\ri j \hbar$, one can rewrite the first equation in \eqref{eq:Baxter-N=2} as
\be
Q_{j+1}+Q_{j-1}=T_j Q_j,
\label{eq:Harper}
\ee
where
\be
Q_j=Q(x+\ri j \hbar),\qquad
T_j=T(x+ \ri j \hbar).
\label{eq:shift1}
\ee
This equation can be also rewritten as the matrix form:
\be
\vec2{Q_{j+1}}{Q_j}
= \mat2{T_j}{-1}{1}{0} \vec2{Q_j}{Q_{j-1}}.
\ee
A short manipulation reveals that 
\be
Q_b+Q_{-b}= \Tr \left[  \mat2{T_{b-1} }{-1}{1}{0} \cdots  \mat2{T_1}{-1}{1}{0}  \mat2{T_0}{-1}{1}{0} \right] Q_0.
\ee
The completely same argument holds for the dual equation in \eqref{eq:Baxter-N=2}.
Thus we have
\be
\wt{Q}_a+\wt{Q}_{-a}= \Tr \left[  \mat2{\wt{T}_{a-1} }{-1}{1}{0} \cdots  \mat2{\wt{T}_1}{-1}{1}{0}  \mat2{\wt{T}_0}{-1}{1}{0} \right] \wt{Q}_0,
\ee
where
\be
\wt{Q}_j=Q(x+2 \pi \ri j),\qquad \wt{T}_j=\wt{T}(x+2\pi \ri j).
\label{eq:shift2}
\ee
Since we have $Q_b=Q(x+2\pi \ri a)=\wt{Q}_a$ for \eqref{eq:hbar-cond}, 
we arrive at the relation
\be
\ba
&\Tr \left[  \mat2{\wt{T}_{a-1} }{-1}{1}{0} \cdots  \mat2{\wt{T}_1}{-1}{1}{0}  \mat2{\wt{T}_0}{-1}{1}{0} \right] \\
&=\Tr \left[  \mat2{T_{b-1} }{-1}{1}{0} \cdots  \mat2{T_1}{-1}{1}{0}  \mat2{T_0}{-1}{1}{0} \right] .
\ea
\label{eq:E-dual-N=2}
\ee
where we used $Q_0=\wt{Q}_0=Q(x)$.
This equation relates $\wt{E}$ to $E$.
To understand this fact more clearly, let us see an example.
We take the particular value $(a,b)=(2,3)$.
It is easy to see
\be
\ba
\Tr \left[  \mat2{T_{2} }{-1}{1}{0} \mat2{T_1}{-1}{1}{0}  \mat2{T_0}{-1}{1}{0} \right]
&=R^{-6} [ E(E^2-3-3R^4)-2\cosh 3x ], \\
\Tr \left[  \mat2{\wt{T}_1}{-1}{1}{0}  \mat2{\wt{T}_0}{-1}{1}{0} \right]
&=\wt{R}^{-4} [ \wt{E}^2-2-2\wt{R}^4-2\cosh 3x ].
\ea
\ee
Since we have $\wt{R}=R^{3/2}$ for $(a,b)=(2,3)$, the $x$-dependence is in precise agreement.
Comparing the $x$-independent term, we find the algebraic relation
\be
\wt{E}^2-2-2\wt{R}^4=E(E^2-3-3R^4),\qquad \hbar=\frac{4\pi}{3},\quad \wt{\hbar}=3\pi.
\ee 
One test of this relation is to compare the discrete spectra of the Hamiltonians \eqref{eq:H-N=2}
and \eqref{eq:H-dual-N=2}, directly.
As explained in \cite{Huang:2014eha}, this can be done by expanding the eigenfunction $Q(x)$
in the orthogonal basis in the Hilbert space $L^2(\mathbb{R})$.
A natural candidate of such a basis is the eigenfunctions for the harmonic oscillator.
In Table~\ref{tab:spectrum-N=2}, we show the first five eigenvalues of the Hamiltonian \eqref{eq:H-N=2} for $\hbar=4\pi/3$
and $\hbar=3\pi$ in the case of $R=1$.
One can check that these eigenvalues indeed satisfy the relation $\wt{E}_n^2-4=E_n(E_n^2-6)$
for each quantum number $n=0,1,2,\dots$.
The same kind of test is possible for given $\hbar$ and $R$.
 
\begin{table}[tb]
\caption{The first five eigenvalues of the Hamiltonian \eqref{eq:H-N=2} for $\hbar=4\pi/3$ and for $\hbar=3\pi$ with $R=1$.
It turns out that these eigenvalues satisfy the non-trivial relation $E_n(3\pi)^2-4=E_n(4\pi/3)(E_n(4\pi/3)^2-6)$
for any non-negative integer $n$.}
\begin{center}
  \begin{tabular}{ccc}
\hline
Eigenvalues & $\hbar=4\pi/3$ & $\hbar=3\pi$ \\
\hline
$E_0$ & $11.038588121924404944$ &  $35.816548625048475896$ \\
$E_1$ & $33.165572067706303312$ &  $190.48792362943094504$ \\
$E_2$ & $76.646795079907244305$ &  $670.68877831711410310$ \\
$E_3$ & $154.53804300167833305$ &  $1920.8735917517111079$ \\
$E_4$ & $285.87088224409482661$ &  $4833.2468516231114653$\\
\hline
  \end{tabular}
\end{center}
\label{tab:spectrum-N=2}
\end{table}

From the practical point of view, it is sufficient to set $x=0$ in \eqref{eq:E-dual-N=2}.
In this case, the equation \eqref{eq:Harper} can be regarded as the Harper equation. 
Let us define a polynomial $P_{a/b}(E, R )$ with degree $b$ by
\be
P_{a/b}(E, R) := \Tr [  A_{b-1}(a/b;R) \cdots A_1(a/b;R)A_0(a/b;R) ] +2.
\label{eq:P-def}
\ee
where
\be
A_k(\tau;R)=\begin{pmatrix} E-2\cos (2\pi k \tau) \quad & -R^2 \\ R^2 & 0 \end{pmatrix}.
\ee 
It turns out that the relation \eqref{eq:E-dual-N=2} at $x=0$ is equivalent to
the condition
\be
P_{b/a}(\wt{E}, \wt{R})=P_{a/b}(E,R).
\label{eq:P-Sdual}
\ee
This is the main result in this section, and provides 
the exact relation between the energy $E$ for $\hbar=2\pi a/b$ and $\wt{E}$ for $\wt{\hbar}=2\pi b/a$.
Some explicit forms of $P_{a/b}(E,R)$ are as follows
\be
\ba
P_{a/1}(E,R)&=E, \\
P_{a/2}(E,R)&=E^2-2(1+(-1)^a)E+2(1-R^4+2(-1)^a) ,\\
P_{1/3}(E,R)&=P_{2/3}(E,R)=E(E^2-3-3R^4), \\
P_{1/4}(E,R)&=P_{3/4}(E,R)=E^4-4(1+R^4)E^2+2(1+R^8), \\
P_{2/4}(E,R)&=E^4-4(2+R^4)E^2+2(9+8R^4+R^8).
\ea
\ee
Since the matrix $A_k(\tau;R)$ is invariant under $\tau \to \tau+1$, 
the polynomial $P_{a/b}(E,R)$ has the following property
\be
P_{|a\pm b|/b}(E,R)=P_{a/b}(E,R).
\ee

\section{Quantum geometry and Hofstadter's butterfly}\label{sec:hof}
In this section, we see the analytic property of a quantum deformed special geometry in
the Calabi--Yau threefold, the local $\mathbb{P}^1 \times \mathbb{P}^1$.
This geometry is important both in gauge theories and in integrable systems.
On one hand, it describes the low energy effective theory of the five-dimensional $\cN{=}1$ pure 
SU(2) super Yang--Mills theory on $\mathbb{R}^4 \times S^1$ via the geometric engineering \cite{Katz:1996fh}.
On the other hand, it is related to the two-particle relativistic Toda lattice \cite{Nekrasov:1996cz}, just reviewed
in the previous section.
In particular, the exact spectrum of the $N=2$ relativistic Toda lattice is
determined by the topological string on this geometry.
We will here reveal that the quantum geometry in the local $\mathbb{P}^1 \times \mathbb{P}^1$
also has a remarkable connection with condensed matter physics.

\subsection{Quantum geometry in the local $\mathbb{P}^1 \times \mathbb{P}^1$}
Let us start by seeing the relation between the quantum geometry in the local $\mathbb{P}^1 \times \mathbb{P}^1$
and the relativistic Toda lattice.
The key concept is local mirror symmetry.
Mirror symmetry states that a Calabi--Yau (CY) manifold has its mirror dual.
The K\"ahler structure of the original CY is mapped to the complex structure of the mirror CY, and vice versa.
In our case, the mirror CY to local $\mathbb{P}^1 \times \mathbb{P}^1$
is described by the following equation, called the mirror curve,
\be
\re^x+z_1 \re^{-x}+\re^p+z_2 \re^{-p}=1,
\label{eq:mirror-curve}
\ee
where $z_1$ and $z_2$ are the complex moduli of the mirror CY.
The mirror curve has enough information to construct the all-genus free energy of
the topological string theory, called the B-model \cite{Marino:2006hs, Bouchard:2007ys}.
Moreover the interesting geometric feature appears when one considers the \emph{quantization}
of the mirror curve.
For our purpose, it is more convenient to shift the variables as
\be
x \to x-\log E, \qquad p \to p+2\log R-\log E,
\ee
and to set
\be
z_1=\frac{1}{E^2}, \qquad z_2=\frac{R^4}{E^2}.
\label{eq:complex-moduli}
\ee
Then the mirror curve \eqref{eq:mirror-curve} is rewritten as 
\be
R^2(\re^{p}+\re^{-p})+\re^{x}+\re^{-x} = E.
\label{eq:mirror-curve-2}
\ee
This is the same form as the Hamiltonian \eqref{eq:H-N=2} of the $N=2$ relativistic Toda lattice.
Now we quantize the variables $x$ and $p$ by $[x,p]=\ri \hbar$.
Since one can write the momentum operator as $p=-\ri \hbar \pd_x$,
the mirror curve \eqref{eq:mirror-curve-2} naturally leads to a difference equation,
which is exactly the Baxter equation in \eqref{eq:Baxter-N=2}.
We conclude that the quantized mirror curve for the local $\mathbb{P}^1 \times \mathbb{P}^1$ 
is related to the quantum eigenvalue problem of the relativistic Toda lattice with just two particles.%
\footnote{In this paper, we focus  only on the case that $x \in \mathbb{R}$ and $\hbar \in \mathbb{R}$.
In principle, one can consider the problem for $x \in \mathbb{C}$ or $\hbar \in \mathbb{C}$, as studied in \cite{Kashani-Poor:2016edc, Krefl:2016svj} for instance.
Though we do not yet see  a visible structure in the general case,
it might give a clue to unify the two spectral problems in the relativistic Toda lattice and in the Hofstadter model. 
}

The main achievement in a series of works \cite{Kallen:2013qla, Huang:2014eha, Grassi:2014zfa, Wang:2015wdy} is that the eigenvalue problems associated
with quantized mirror curves are completely determined by exact quantization conditions in terms of the topological
strings on the corresponding geometries.
For the local $\eP^1 \times \eP^1$, the quantization condition is
\be
\frac{\pd}{\pd t} F(t,t-\log R^4;q)+\frac{\pd}{\pd \wt{t}}\, F(\wt{t}, \wt{t}-\log \wt{R}^4; \wt{q} )
=2\pi \( n+\frac{1}{2} \),
\qquad n \in \mathbb{Z}_{\geq 0}.
\label{eq:EQC}
\ee
We need to explain the notation of this equation.
The function $F(t_1, t_2;q)$ is related to the free energy of the refined topological string in the Nekrasov--Shatashvili limit.
It has two K\"ahler moduli $t_1$ and $t_2$, which parametrize the size of two $\eP^1$'s, with the parametrization $t_1=t$ and $t_2=t-\log R^4$.
It takes the following form
\be
F(t,t-\log R^4;q)=\frac{t^3}{6\hbar}-\frac{\log R^4}{4\hbar} t^2-\( \frac{\pi}{3\hbar}+\frac{\hbar}{12} \)t
+2F_\text{NS}(t,t-\log R^4;q),
\ee
where $F_\text{NS}(t_1,t_2;q)$ is the Nekrasov--Shatashvili free energy for local $\eP^1 \times \eP^1$,
whose explicit form is given by (see \cite{Huang:2014nwa} for example)
\be
\ba
F_\text{NS}(t_1,t_2;q)=-\sum_{j_L,j_R}\sum_{w,d_j \geq1} \frac{1}{2w^2} N^{d_1,d_2}_{j_L,j_R}
\frac{\sin \frac{\hbar w}{2}(2j_L+1) \sin \frac{\hbar w}{2} (2j_R+1)}{\sin^3 \frac{\hbar w}{2}} \re^{-w (d_1 t_1+d_2 t_2)}.
\ea
\label{eq:F-NS}
\ee
In the above sum, $j_R$ and $j_L$ run for $0,1/2,1,3/2,\dots$.
The integers $N^{d_1,d_2}_{j_L,j_R}$ are  called the refined BPS invariants, and encode the geometrical information on the local $\eP^1 \times \eP^1$.
Their explicit values are found in \cite{Iqbal:2007ii}.
Using these data, the very first few terms are given by
\be
\ba
\ri F_\text{NS}(t_1,t_2;q)=\frac{q+1}{q-1}(\re^{-t_1}+\re^{-t_2})+\frac{q^2+1}{4(q^2-1)}(\re^{-2t_1}+\re^{-2t_2}) \\
+\frac{(q+1)^2(q^2+1)}{q(q^2-1)}\re^{-t_1-t_2}+\cdots.
\ea
\label{eq:F-NS-2}
\ee
The dual variables in \eqref{eq:EQC} are defined by
\be
\wt{t}=\frac{2\pi t}{\hbar},\qquad
\log \wt{R}=\frac{2\pi}{\hbar} \log R,\qquad
\wt{q}=\re^{\ri \wt{\hbar}},\qquad
\wt{\hbar}=\frac{4\pi^2}{\hbar}.
\ee
These just correspond to the modular dual transform in the relativistic Toda lattice.

Let us remark on the quantization condition \eqref{eq:EQC}.
It is obvious to see that the Nekrasov--Shatashvili free energy \eqref{eq:F-NS} or \eqref{eq:F-NS-2}
has an infinite number of poles at $\hbar=2\pi a/b$ ($a,b \in \mathbb{Z}$).
However, these poles are precisely cancelled by the modular dual part, i.e.,
the second term on the left hand side in \eqref{eq:EQC}.
This cancellation mechanism was first found in ABJM theory \cite{Hatsuda:2012dt, Hatsuda:2013oxa}.
From the viewpoint of quantum geometry, the Nekrasov--Shatashvili free energy
corresponds to the quantum B-period \cite{Aganagic:2011mi}.
In this sense, the quantum B-period itself is ill-defined for $\hbar=2\pi a/b$,
but the combination with its modular dual gives a well-defined function on the whole real line of $\hbar$.%
\footnote{This structure is widely found in functions, e.g. the non-compact quantum dilogarithm, 
that have the modular double property. See subsection 5.4.2 in \cite{Meneghelli:2015sra}, for example.}
Moreover, the free energy $F_\text{NS}(t,t-\log R^4;q)$ is an expansion in terms of $\re^{-t}$,
and thus its modular dual is an expansion in terms of $\re^{-\wt{t}}=\re^{-2\pi t/\hbar}$, which is non-perturbative in $\hbar$.
Therefore, the modular dual part is not visible in the semiclassical analysis $\hbar \to 0$.

The quantization condition determines a discrete value of $t$ for a given quantum number $n$.
To know the eigenvalues of the Hamiltonian, we need a precise relation between $t$ and $E$.
This can be done by the so-called quantum A-period:
\be
-t=\log z+ \wt{\Pi}_A(z, R^4 z; q),\qquad
z=\frac{1}{E^2}.
\ee
Inverting this relation, one can recover the eigenvalue $E$.
The quantum A-period around $z=0$ can be computed from the difference equation \eqref{eq:Baxter-N=2}, as explained in \cite{Aganagic:2011mi}.
We review it in Appendix~\ref{sec:period}.
The first few terms are given by
\be
\ba
\wt{\Pi}_A(z, R^4 z; q)&=2(1+R^4)z+\(3+8R^4+3R^8+2R^4(q+q^{-1}) \)z^2 \\
&\quad+\biggl( \frac{20}{3}+32R^4+32 R^8+\frac{20R^{12}}{3}
+12R^4(1+R^4)(q+q^{-1}) \\
&\quad+2R^4(1+R^4)(q^2+q^{-2}) \biggr)z^3
+\cO(z^4).
\ea
\label{eq:quantum-A}
\ee

The important consequence of the quantization condition \eqref{eq:EQC}
is that it is obviously invariant under the modular dual transform
\be
(t,R,q) \leftrightarrow (\wt{t}, \wt{R}, \wt{q}).
\ee
As already mentioned in \cite{Hatsuda:2015qzx}, this remarkable invariance is understood as 
a consequence of the modular double in the relativistic Toda lattice.
In particular, the relation between $t$ and $\wt{t}$ implicitly relates $E$ to $\wt{E}$:
\be
\wt{t}(\wt{E}, \wt{R};\wt{q})=\frac{2\pi}{\hbar} t(E, R;q).
\label{eq:t-S-trans}
\ee
In fact, one can check, by using \eqref{eq:quantum-A}, that this relation gives the equivalent relation to \eqref{eq:P-Sdual}
for $\hbar=2\pi a/b$, or conversely, by using \eqref{eq:P-Sdual} and \eqref{eq:quantum-A}, one can confirm the relation \eqref{eq:t-S-trans}.
This fact provides further (indirect) evidence of the validity of the quantization condition \eqref{eq:EQC}.

\subsection{Quantum flat coordinates}
In this subsection, we investigate the analytic property of the flat coordinate $t$ in the local $\eP^1 \times \eP^1$.
As seen in \eqref{eq:quantum-A}, it receives quantum corrections.
It was observed in \cite{Hatsuda:2013oxa} that this expansion is a convergent series for $|q|=1$.%
\footnote{For $|q|=1$, the large order behavior of the coefficients of $z^n$ shows $\exp(C n)$ with a constant $C$,
while for $|q| \ne 1$, the large order behavior seems to be  $\exp(C n^2)$ \cite{Hatsuda:2013oxa}.}
However, it seems technically difficult to resum it for general $q$.
Surprisingly, as shown here, we can perform the resummation  with a trick, if $q$ is a root of unity.
The resulting analytic property of $t$ turns out to have a very rich structure.

We first observe that the coefficients of the small $z$-expansion of $\wt{\Pi}_A(z,R^4 z;q)$ are Laurent polynomials of $q$.
Also these polynomial are symmetric under the exchange of $q$ and $q^{-1}$.
We confirmed these observations up to order $z^9$.
This implies the symmetries
\be
\wt{\Pi}_A(z,R^4 z;\re^{2\pi \ri} q)=\wt{\Pi}_A(z,R^4 z;q),\qquad
\wt{\Pi}_A(z,R^4 z;q^{-1})=\wt{\Pi}_A(z,R^4 z;q).
\label{eq:A-symmetry}
\ee
The former corresponds to the shift $\hbar \to \hbar +2\pi$, while the latter to the reflection $\hbar \to -\hbar$.
We assume that these symmetries exactly work for any complex $z$.
Using these symmetries and the S-dual transform \eqref{eq:t-S-trans}, 
we can compute the exact form of the flat coordinate for $\hbar=2\pi \tau$ where $\tau=a/b$ with coprime integers $a$ and $b$, using Euclid's algorithm.
For concreteness, let us consider an example: $\tau=2/5$. In this case, the dual modulus is $\tau= 5/2$.
Using the shift symmetry \eqref{eq:A-symmetry}, the quantum A-period at $\tau=5/2$ is equal to that at $\tau=1/2$.
We then use the S-dual transform again, and obtain the modulus $\tau=2$.
Of course, the A-period at this value is equivalent to that at $\tau=1$.
In this way, the computation for $\tau=2/5$ is mapped into that for $\tau=1$. 
The basic flow this reduction is summarized as
\be
\frac{2}{5} \stackrel{S}{\to} \frac{5}{2} \stackrel{T}{\to} \frac{1}{2} \stackrel{S}{\to} 2 \stackrel{T}{\to} 1,
\ee
where $S$ signifies the S-transform $\tau \to 1/\tau$, while $T$ stands for the translation $\tau \to \tau -1$.
In order to relate the flat coordinate $t$ at $\tau=2/5$ to that at $\tau=1$,
one has to use the chain of the transforms carefully.
Taking into account the translation invariance \eqref{eq:A-symmetry}, it is easy to see
\be
\ba
t(E, R ;\re^{2\pi \ri \cdot 2/5})&=\frac{2}{5}t'(E', R';\re^{2\pi \ri \cdot 5/2})
=\frac{2}{5}t'(E', R';\re^{2\pi \ri \cdot 1/2})\\
&=\frac{2}{5}\cdot \frac{1}{2} \wt{t}(\wt{E}, \wt{R}; \re^{2\pi \ri \cdot 2})
=\frac{1}{5} \wt{t}(\wt{E}, \wt{R}; 1),
\ea
\ee 
where by using \eqref{eq:P-Sdual}, each energy is also related to
\be
P_{2/5}(E,R)=P_{5/2}(E', R')=P_{1/2}(E', R')=P_{2/1}(\wt{E}, \wt{R})=\wt{E},
\ee
and we have $\wt{R}=(R')^2=R^5$.
We conclude that the flat coordinate for $q=\re^{4\pi \ri/5}$ is exactly given by
\be
t(E, R; q=\re^{4\pi \ri/5})=\frac{1}{5} \wt{t}(\wt{E}, \wt{R}; \wt{q}=1), \qquad
\wt{E}=P_{2/5}(E, R), \quad \wt{R}=R^5,
\ee
where the polynomial $P_{2/5}(E,R )$ is explicitly computed by the formula \eqref{eq:P-def}.
Our remaining task is to evaluate the A-period at $q=1$.
This seems difficult for general $R$,
but as computed in Appendix~\ref{sec:period}, we can express its derivative with respect to $E$ in closed form.
Using \eqref{eq:classical-A}, we thus find
\be
\frac{\pd t(E,R ;\re^{4\pi \ri/5})}{\pd E}
=\frac{2}{5\pi}\frac{\pd \wt{E}}{\pd E} \biggl( \frac{\wt{E}^2}{4}-(1-\wt{R}^2)^2 \biggr)^{-1/2} \eK \( \frac{16\wt{R}^2}{\wt{E}^2-4(1-\wt{R}^2)^2} \),
\label{eq:t-exact-2/5}
\ee
where $\eK(z)$ is the complete elliptic integral of the first kind. Our convention of the elliptic integral is
\be
\eK(z):=\int_{0}^{\pi/2} \frac{\rd \phi}{\sqrt{1-z \sin^2 \phi}}.
\label{eq:ellipticK}
\ee
Let us test this result.
From the exact result \eqref{eq:t-exact-2/5}, one easily obtains the following large $E$-expansion:
\be
\ba
\frac{\pd t(E,R ;\re^{4\pi \ri/5})}{\pd E} &= \frac{2}{E }+\frac{4 (R^4+1)}{E^3}
+\frac{4 (3 R^8-(\sqrt{5}-7 ) R^4+3)}{E^5} \\
&\quad+\frac{10 (R^4+1) (4 R^8+(11-3 \sqrt{5}) R^4+4 )}{E^7}+\cO(1/E^{9}).
\ea
\ee
Completely the same expansion is obtained from the small $z$-expansion \eqref{eq:quantum-A} by setting $q=\re^{4\pi \ri/5}$.
We confirmed this agreement up to order $1/E^{19}$.

Since any rational number $\tau=a/b$ can be reduced to $\tau=1$ by repeating the S-transform and the T-transform,
the above result is easily generalized to arbitrary $\tau=a/b$.
We finally find
\be
\frac{\pd t(E,R ;q=\re^{2\pi \ri a/b})}{\pd E}
=\frac{P'_{a/b}(E, R)}{\pi b R^b} \frac{\eK(1/F)}{\sqrt{F}},
\label{eq:t-exact}
\ee
where
\be
F=\frac{P_{a/b}(E,R)^2-4(1-R^{2b})^2}{16R^{2b}}.
\ee
This is one of the main results in this section.
We can evaluate the quantum flat coordinate $t$ whenever $q$ is a root of unity!
For $R=1$, the result can be further simplified.
In this case, we find
\be
\ba
t(E,1; \re^{2\pi \ri a/b})=-\frac{1}{b} \left[ \log \wt{z}+4\wt{z} \,{}_4F_3 \( 1,1, \frac{3}{2},\frac{3}{2};2,2,2;16\wt{z} \) \right],\quad
\wt{z}=\frac{1}{P_{a/b}(E,1)^2}.
\label{eq:t-exact-R=1}
\ea
\ee
In the limit $R \to 0$, the result is drastically simplified.
In this limit, the quantum A-period does not depend on $q$, and it is always the
same as that at $q=1$.
Using \eqref{eq:Pi-A-LR}, one finds
\be
t(E,0;q)=2 \log \biggl[ \frac{E+\sqrt{E^2-4}}{2} \biggr].
\ee

\begin{figure}[tb]
\begin{center}
\begin{tabular}{cc}
\hspace{-3mm}
\includegraphics[width=.45\textwidth]{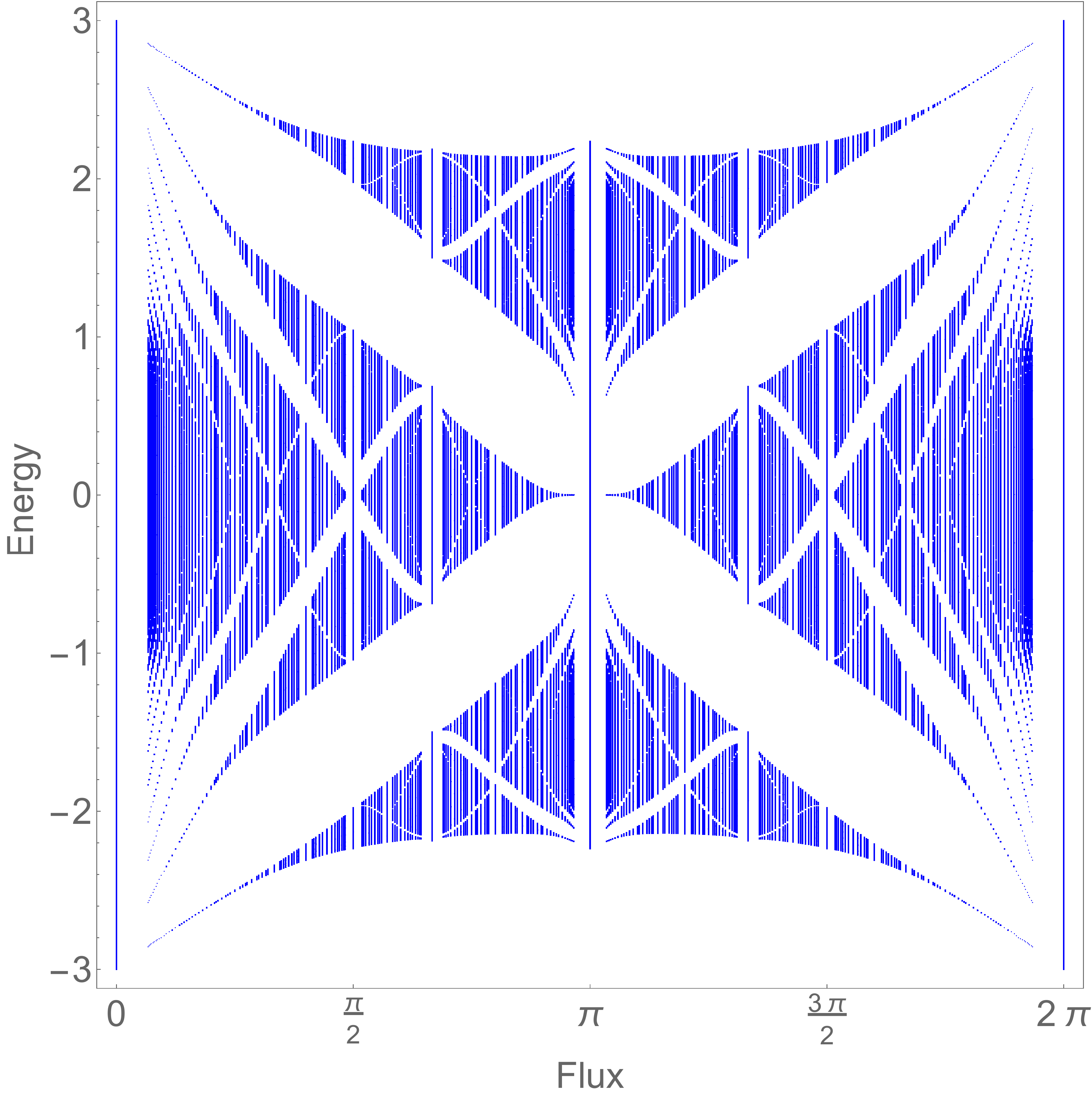}
\hspace{6mm}
\includegraphics[width=.45\textwidth]{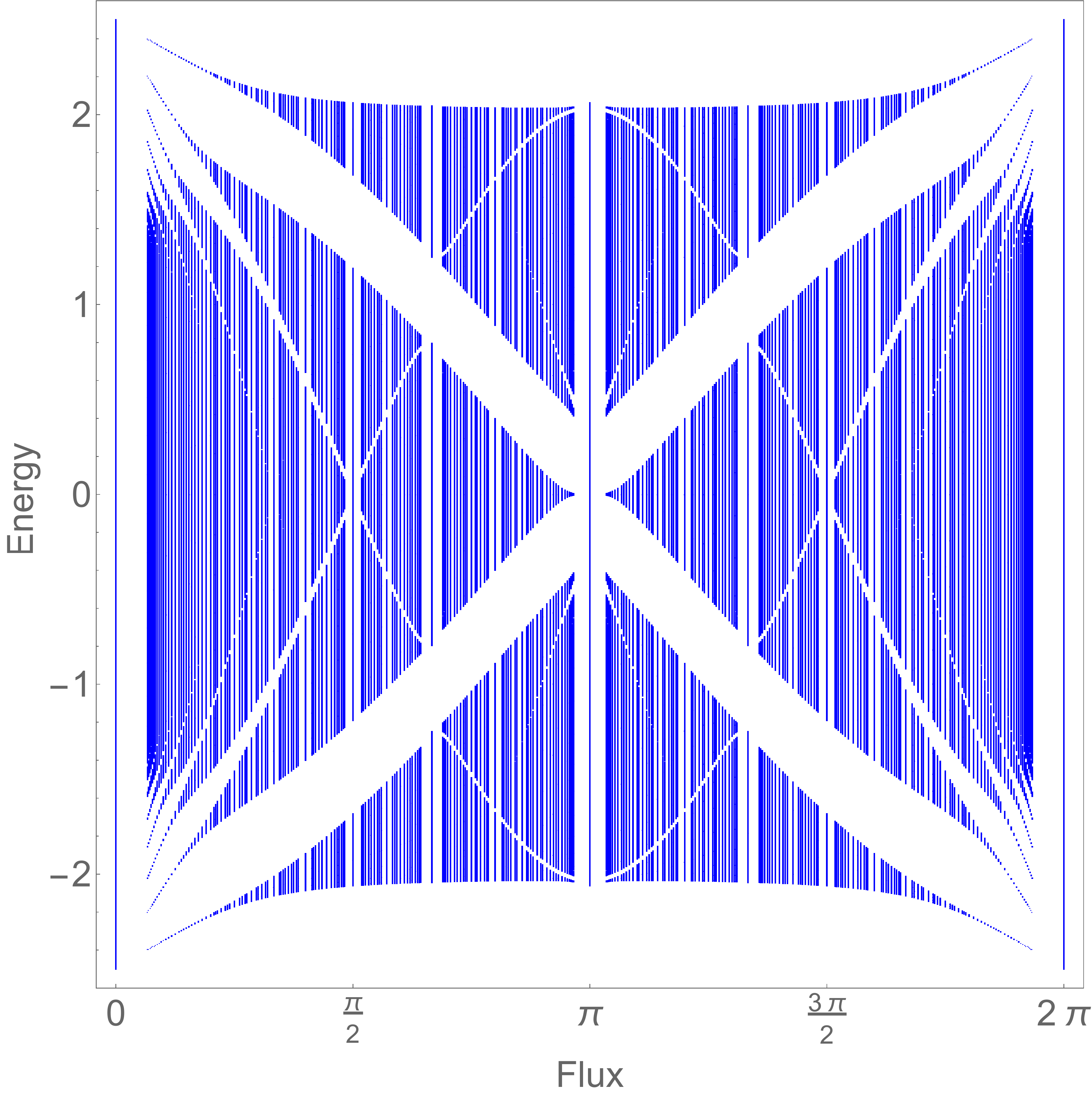}
\vspace{-5mm}
\end{tabular}
\end{center}
  \caption{The branch cut structures for $R^2=1/2$ (Left) and for $R^2=1/4$ (Right).}
  \label{fig:butterfly-aniso}
\end{figure}

Let us proceed to the study of the analytic property of the flat coordinate.
Without loss of generality, we can assume $R \leq 1$.
The complete elliptic integral $\eK(1/F)$ has a branch cut along $1/F \geq 1$, i.e.,
\be
\frac{16 R^{2b}}{P_{a/b}(E,R)^2-4(1-R^{2b})^2} \geq 1.
\ee
This leads to the condition
\be
2(1-R^{2b}) \leq |P_{a/b}(E,R)|  \leq 2(1+R^{2b}).
\label{eq:branch-1}
\ee
Also, the factor $\sqrt{F}$ has branch cuts along $F \leq 0$, and this leads to
\be
|P_{a/b}(E,R)| \leq 2(1-R^{2b}).
\label{eq:branch-2}
\ee
Combining these, we conclude that the function \eqref{eq:t-exact} has branch cuts along
\be
|P_{a/b}(E,R)| \leq 2(1+R^{2b}).
\label{eq:branch}
\ee
It is observed that all the branch cuts are on the real axis in the complex energy plane, and the number of cuts is at most $b$.
The branch cut structure determined by this equation shows a quite complicated behavior in the energy plane.
In Fig.~\ref{fig:butterfly}, we show it for $R=1$. 
We plot the branch cuts for $\hbar=2\pi a/b$ with all possible coprime integers $1 \leq a \leq b \leq 30$.
As already mentioned in the introduction, this figure is well-known as Hofstadter's butterfly in the two-dimensional
electron system.
In fact, the same condition as \eqref{eq:branch} for $R=1$ was obtained in \cite{Hof},
though its derivation looks quite different. 
We also show the case of $R \ne 1$ in Fig.~\ref{fig:butterfly-aniso}.
The left figure is for $R^2=1/2$, while the right for $R^2=1/4$.
They correspond to anisotropic cases in the Hofstadter problem.
We conclude that the branch cuts of $t$ in the energy plane precisely correspond to the energy bands in the Hofstadter model.

If $E$ satisfies the condition \eqref{eq:branch}, the flat coordinate $t$ takes complex values.
In this regime, it is more convenient to use another expression.
Using the identity for the complete elliptic integral:
\be
\eK(1/z)=\sqrt{z} [ \eK(z)+\ri \eK(1-z) ],
\label{eq:K-inverse}
\ee
we obtain
\be
\frac{\pd t(E,R ;q=\re^{2\pi \ri a/b})}{\pd E}
=\frac{P'_{a/b}(E, R)}{\pi b R^b} [ \eK(F)+\ri \eK(1-F)].
\label{eq:t-exact-2}
\ee
In the next subsection, we will see that the imaginary part of this equation
also has a nice physical interpretation in the 2d electron system.
In the case of \eqref{eq:branch-1}, it is easy to see
\be
\im \biggl[ \frac{\pd t}{\pd E} \biggr] 
=\frac{P'_{a/b}(E, R)}{\pi b R^b} \eK(1-F),\quad
\bigl( 2(1-R^{2b}) \leq |P_{a/b}(E,R)|  \leq 2(1+R^{2b}) \bigr).
\label{eq:Im-t-1}
\ee
In the case of \eqref{eq:branch-2}, $\eK(F)$ is still real, but $\eK(1-F)$ takes complex values.
Using \eqref{eq:K-inverse} again, one finally finds
\be
\im \biggl[ \frac{\pd t}{\pd E} \biggr] 
=\frac{P'_{a/b}(E, R)}{\pi b R^b}\frac{1}{\sqrt{1-F}} \eK\( \frac{1}{1-F} \), \qquad
\bigl( |P_{a/b}(E,R)| \leq 2(1-R^{2b}) \bigr).
\label{eq:Im-t-2}
\ee

\subsection{Comparing to two-dimensional electrons in a magnetic field}\label{subsec:2d-electron}
In this subsection, we review the analysis of the 2d electron system
with a periodic potential in a uniform magnetic field,
and compare the result with the one obtained in the previous subsection.

If the effect of the magnetic field is sufficiently smaller than the potential,
we can use the tight-binding approximation.
We here consider the 2d electrons on the square lattice with lattice spacing $a=1$.
The tight-binding Hamiltonian is
\be
H=T_x+T_x^\dagger+\lambda(T_y +T_y^\dagger),
\label{eq:H-TB}
\ee
where $T_x$ and $T_y$ are magnetic translation operators.
They satisfy the following algebraic relations \cite{WZ, NPB_Wiegmann_Zabrodin, Faddeev:1993uk}
\be
\ba
T_x T_x^\dagger = T_y T_y^\dagger =1, \qquad
T_x T_y= q T_y T_x,
\ea
\label{eq:Tx-Ty}
\ee
where $q=\re^{\ri \phi}$, with $\phi$ being a flux through an elementary plaquette, and $\lambda$ is the parameter that describes the anisotropy between the hopping amplitudes in the $x$ and $y$ directions.
If the magnetic field is turned off, the dispersion relation is simply given by
\be
E=2\cos k_x+2\lambda \cos k_y,
\ee
where $(k_x, k_y)$ is the wave vector.
When the magnetic field is turned on, the translation operators $T_x$ and $T_y$ no longer
commute, as in \eqref{eq:Tx-Ty}.
In this picture, one can elevate the dispersion relation to the Peierls--Onsager effective Hamiltonian \cite{Kohmoto1}
\be
H=2\cos \Pi_x+2 \lambda \cos \Pi_y,\qquad [\Pi_x, \Pi_y]=\ri \phi.
\label{eq:H-Hof}
\ee
In fact, the relation \eqref{eq:Tx-Ty} is satisfied by setting $T_x=\re^{\ri \Pi_x}$ and $T_y=\re^{\ri \Pi_y}$. 
If we rename $\Pi_x \to x$, $\Pi_y \to p$ and $\phi \to \hbar$, the Hamiltonian is
just the same one in \eqref{eq:harper}.
It turned out that the Hamiltonian \eqref{eq:H-Hof} indeed has the same spectrum
as the original tight-binding Hamiltonian \eqref{eq:H-TB}.

When the effect of the magnetic field is far larger than the periodic potential, we can project the system to the lowest Landau level (LLL), with a perturbation. 
As explained in \cite{TKNN}, in this case, one gets almost the same Hamiltonian but the flux in the commutation 
relation is flipped as $\phi \to 4\pi^2/\phi$.
We do not consider this case in detail, except for noting that curiously this quantization parameter is exactly the modular dual of the situation above.

Let us return to the tight-binding approximation.
Following the argument of Hofstadter \cite{Hof}, the tight-binding Hamiltonian \eqref{eq:H-TB} leads to the following Harper equation \cite{Kohmoto1, Kohmoto2}:
\be
\re^{\ri k_x} \psi_{n+1}+\re^{-\ri k_x}\psi_{n-1}+2\lambda \cos (k_y'+2\pi n \tau) \psi_n = E \psi_n,
\ee
where $\phi=2\pi \tau$ and we wrote $k_y$ as $k_y'+2\pi n\tau$.
In the following, we consider the case
\be
\tau=\frac{a}{b},
\ee
where $a$ and $b$ are coprime positive integers.
The magnetic Brillouin zone is identified as
\be
0 \leq k_x \leq \frac{2\pi}{b},\qquad
0 \leq k_y' \leq 2\pi.
\label{eq:Brill}
\ee
We also have the boundary condition $\psi_{n+b}=\psi_n$.
The spectrum of the Harper equation is determined by the equation \cite{Kohmoto1, Kohmoto2}
\be
F_{a/b}(E, \lambda)+2(1+\lambda^b)=2\cos(b k_x)+2\lambda^b \cos (b k_y'),
\ee
where $F_{a/b}(E, \lambda)$ is a characteristic polynomial of the form
\be
\ba
F_{a/b}(E, \lambda)=\det
\begin{pmatrix}
M_1(E,\lambda) & -1   & 0   & \cdots & 0 & 0 & -1 \\
-1  &  M_2(E,\lambda) & -1 & \cdots & 0 & 0 & 0 \\
\vdots  & \vdots & \vdots  &  & \vdots & \vdots & \vdots \\
0  &  0  &  0  &  \cdots  &  -1  &  M_{b-1}(E,\lambda)  &  -1  \\
-1  &  0  &  0  &  \cdots  &  0  &  -1  &  M_b(E,\lambda)  
\end{pmatrix},
\ea
\ee
with 
\be
M_n(E, \lambda)=E-2\lambda \cos (2\pi n \tau).
\ee
It turns out that this polynomial is precisely related to $P_{a/b}(E,R)$
\be
P_{a/b}(E,R)=F_{a/b}(E, R^2)+2(1+R^{2b}) \quad
\text{if $a$ and $b$ are coprime}.
\ee
Since $(k_x, k_y')$ takes the values in the magnetic Brillouin zone \eqref{eq:Brill}, we conclude that the energy bands are determined by the condition
\be
|F_{a/b}(E, \lambda)+2(1+\lambda^b)| \leq 2(1+\lambda^{b}).
\ee
This condition is exactly the same as \eqref{eq:branch} with the identification $\lambda=R^2$. 

Next, let us study the density of states.
It is known that the density of states in the 2d electrons with anisotropy has two expressions (see e.g.,  \cite{montroll1956, Kohmoto2}%
\footnote{The isotropic case ($\lambda=1$) was first studied by Wannier,
Obermair, and Ray in \cite{wannier1979}.}).
Let us introduce a short notation:
\be
P(E):=P_{a/b}(E, \sqrt{\lambda})= F_{a/b}(E, \lambda)+2(1+\lambda^b).
\ee
For $2(1-\lambda^{b}) \leq |P(E)|  \leq 2(1+\lambda^{b})$, the density of states is given by
\be
\rho(E)= \frac{P'(E)}{2\pi^2 b \lambda^{b/2}}  \eK \( \frac{4(1+\lambda^b)^2-P(E)^2}{16\lambda^b} \).
\ee
For $|P(E)|  \leq 2(1-\lambda^{b})$, the expression is more complicated,
\be
\rho(E)= \frac{P'(E)}{2\pi^2 b \lambda^{b/2}}\biggl( \frac{16\lambda^b}{4(1+\lambda^b)^2-P(E)^2}  \biggr)^{1/2}
\eK \( \frac{16\lambda^b}{4(1+\lambda^b)^2-P(E)^2} \).
\ee
Now we compare these results with the imaginary part of $\pd t/\pd E$, see \eqref{eq:Im-t-1} and \eqref{eq:Im-t-2}.
One easily see that these are exactly related by
\be
\rho(E)=\frac{1}{2\pi} \im \biggl[ \frac{\pd t(E,R ;q=\re^{2\pi \ri a/b})}{\pd E} \biggr],\qquad
\lambda=R^2.
\ee
As shown in \cite{montroll1956, Kohmoto2}, the density of states exhibits a logarithmic singularity (van Hove singularity) at the middle of each subband.%
\footnote{In the terminology of the CY moduli space, these singularities probably correspond to the orbifold points, while the edges of the energy bands
should correspond to the conifold points.}

\begin{figure}[tb]
\begin{center}
\begin{tabular}{cc}
\hspace{-3mm}
\includegraphics[width=.3\textwidth]{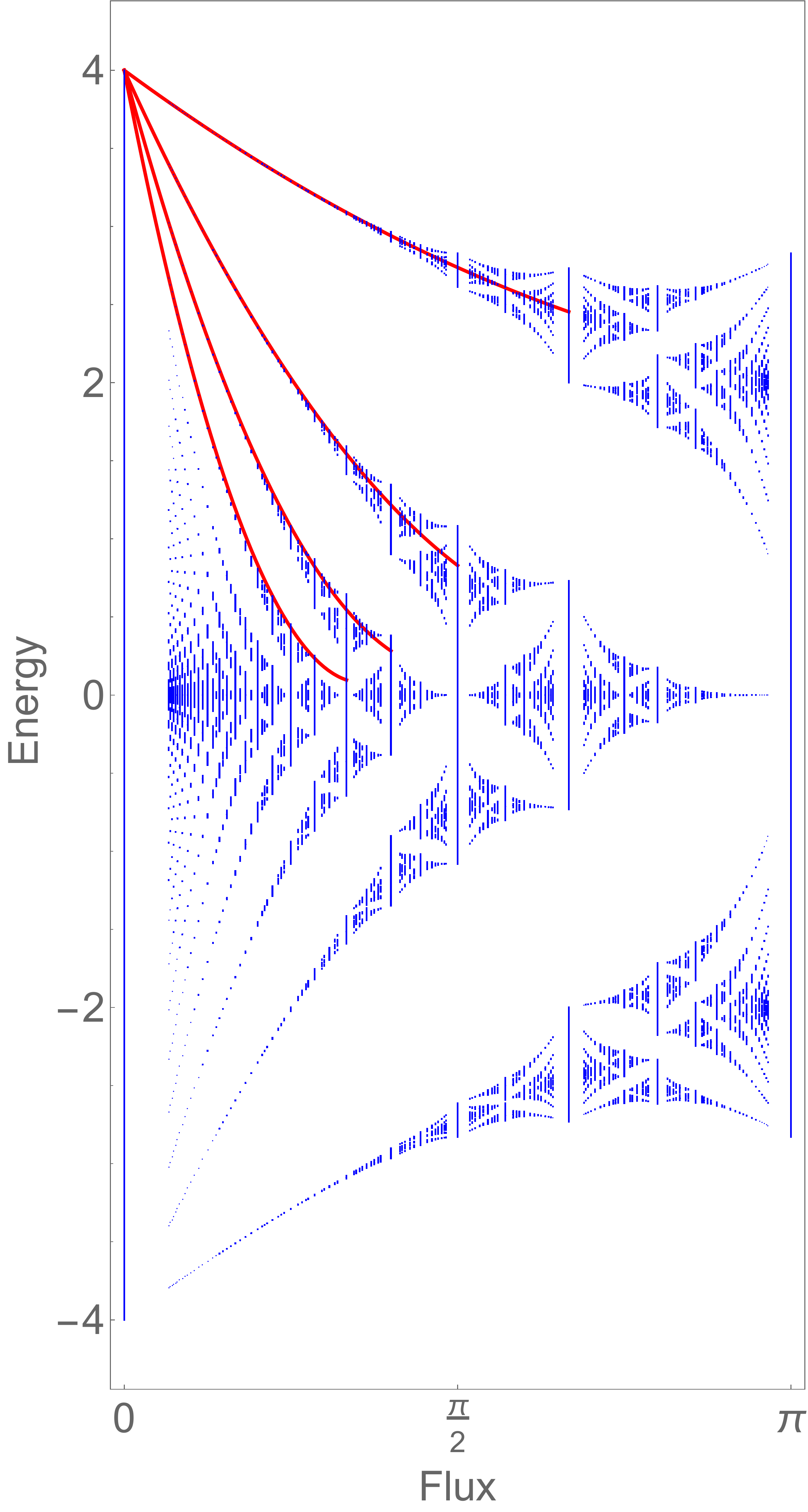}
\hspace{4mm}
\includegraphics[width=.3\textwidth]{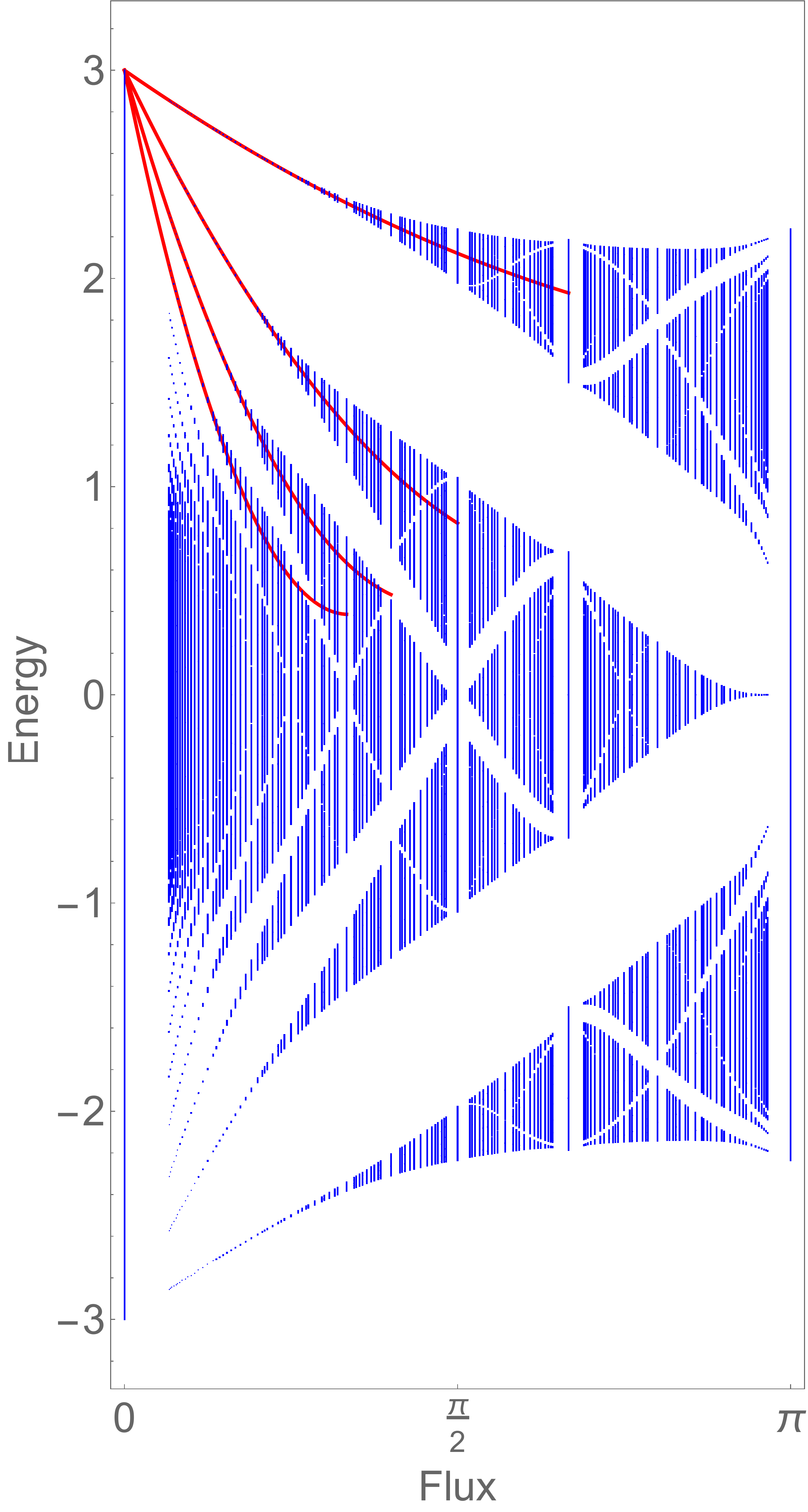}
\hspace{4mm}
\includegraphics[width=.3\textwidth]{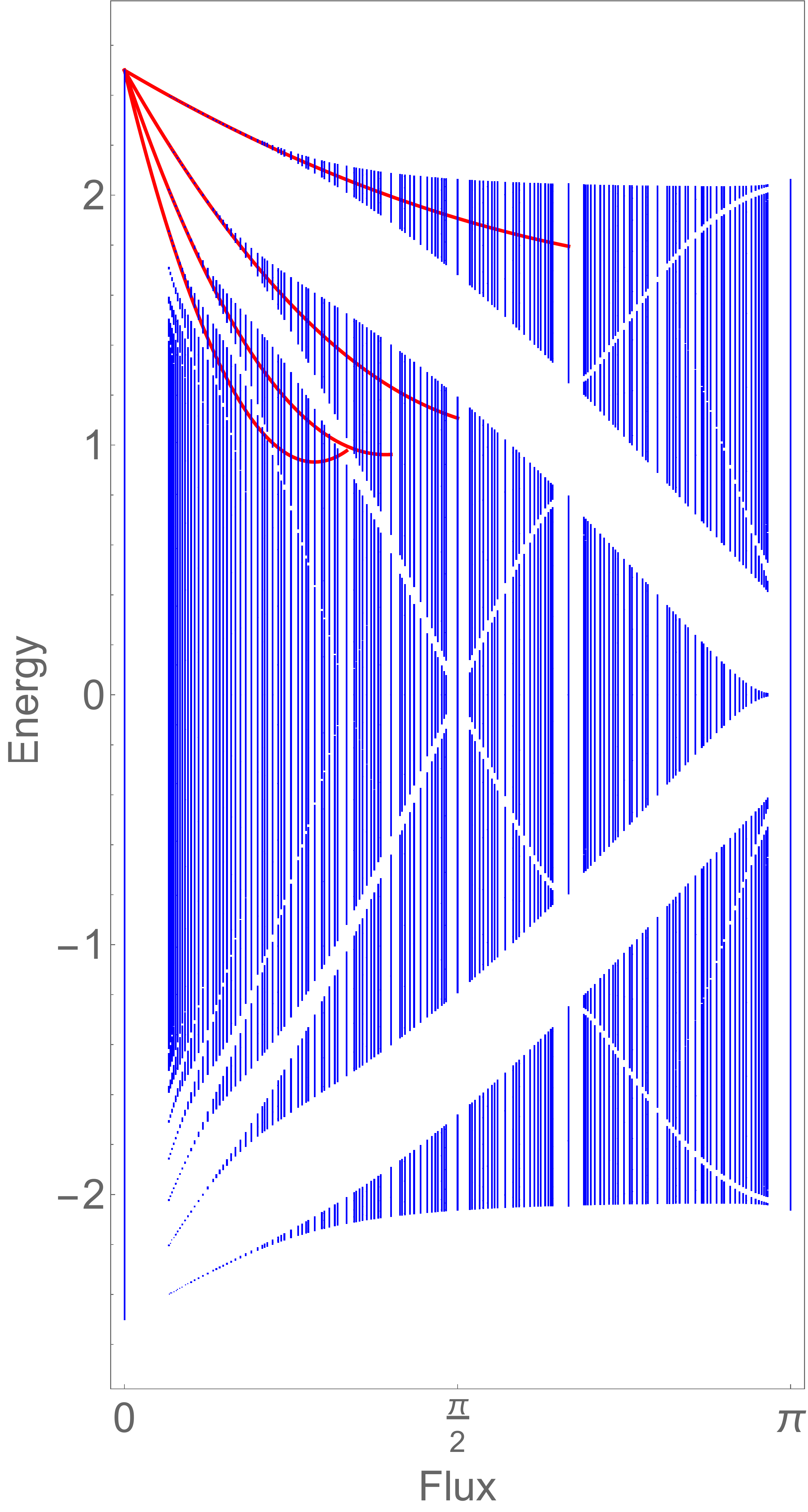}
\vspace{-5mm}
\end{tabular}
\end{center}
  \caption{The weak flux behaviors for $\lambda=1$ (Left), for $\lambda=1/2$ (Middle) and for $\lambda=1/4$ (Right) are shown.
We show the first four graphs ($0 \leq n \leq 3$) of the expansion \eqref{eq:Hof-semi} by the red solid lines.
In all of these case, the semiclassical expansion captures the positions of the bands for $\phi \sim 0$.}
  \label{fig:butterfly-weak}
\end{figure}

Finally we would like to comment on the semiclassical limit.
In the weak magnetic field limit $\phi \to 0$, one can treat the Hamiltonian \eqref{eq:H-Hof} semiclassically.
In this case, the spectrum is located near the extremum $E=2(1+\lambda)$ or $E=-2(1+\lambda)$.
This can be understood by expanding the Hamiltonian \eqref{eq:H-Hof} as
\be
H=2(1+\lambda)-(\Pi_x^2+\lambda \Pi_y^2)+\frac{1}{12}(\Pi_x^4+\lambda \Pi_y^4)+\cdots.
\ee
This can be seen as a perturbation of the harmonic oscillator.
The terms $\Pi_x^{2m}$ and $\Pi_y^{2m}$ give contributions of order $\phi^{m}$.
In the standard perturbation technique, one immediately finds
the following semiclassical expansion
\be
E=2(1+\lambda)-\sqrt{\lambda} (2n+1)\phi +\frac{1+\lambda}{16}(2n^2+2n+1)\phi^2+\cO(\phi^3).
\label{eq:Hof-semi}
\ee
Near $\phi=0$, the width of each band is exponentially narrow, and
the spectrum can be regarded as the Landau levels labelled by $n$ in \eqref{eq:Hof-semi}.
We show the behavior near $\phi=0$ in Fig.~\ref{fig:butterfly-weak}.
The semiclassical expansion indeed explains the position of the bands.

Similarly, if we consider the semiclassical limit of the Hamiltonian \eqref{eq:H-N=2},
we find
\be
E^\text{Toda}=2(1+R^2)+R(2n+1) \hbar+\frac{1+R^2}{16}(2n^2+2n+1)\hbar^2+\cO(\hbar^3).
\label{eq:Toda-semi}
\ee
These two expansions are simply related by the replacement $\phi \to -\hbar$.
This is easily understood since the two Hamiltonians are connected by the analytic continuation
$x \to \ri \Pi_x$ and $p \to \ri \Pi_y$.
Both of the semiclassical expansions above are asymptotic divergent series,
but there is a crucial difference.
As observed in \cite{Hatsuda:2015fxa}, the expansion \eqref{eq:Toda-semi} is an alternating sum.
This means that the Borel transform of \eqref{eq:Toda-semi} does not have any singularities
on the positive real axis, and its Borel sum is well-defined for $\hbar>0$.
On the other hand, the expansion \eqref{eq:Hof-semi} is a non-alternating sum,
and it should have singularities on the positive real axis.
In this case, the Borel sum along the positive real axis is not defined, and
one has to avoid these singularities by deforming the integration contour.
There are choices in how to deform the contour.
This ambiguity is of order $\re^{-1/\phi}$ and must be annihilated by additional
non-perturbative corrections to the semiclassical expansions.
In this case, one needs a trans-series expansion to explain the spectrum
for finite $\phi$.
Roughly speaking, the non-perturbative order $\re^{-1/\phi}$ is also related to the width of the bands,
and thus it is extremely narrow in the weak flux limit.
Recently, the non-perturbative band splitting in the very similar (but different) setup was also confirmed in \cite{Krefl:2016svj}.%
\footnote{In \cite{Krefl:2016svj}, the band splitting was observed in the spectral problem for $H=\re^{p}+\re^{-p}+\lambda(\re^{\ri x}+\re^{-\ri x})$
with $[x,p]=\ri \hbar$ ($x \in \mathbb{R}$, $\hbar>0$, $\lambda >0$).}

To close this section,
we summarize the difference between the $N=2$ relativistic Toda lattice and the Hofstadter model in Table~\ref{tab:com}.
We stress that the local $\eP^1 \times \eP^1$ geometry has the complete spectral information in the both models.

\begin{table}[tb]
\caption{The $N=2$ relativistic Toda lattice vs the Hofstadter model.}
\begin{center}
  \begin{tabular}{c|cc}
\hline
Model & Relativistic Toda & Hofstadter \\
\hline
Energy domain & $E \geq 2(1+R^2)$ & $|E| \leq 2(1+\lambda)$ \\
Spectrum & Discrete &  Finite bands \\
Spectral information & B-period (+ its modular dual) &  A-period \\
Semiclassical expansion& Borel summable &  Non-Borel summable \\
\hline
  \end{tabular}
\end{center}
\label{tab:com}
\end{table}

\section{Conclusions}\label{sec:conclusions}

In this paper, we found Hofstadter's butterfly in the quantum local $\eP^1 \times \eP^1$ geometry.
We made a further study of the recent conjectural solution to the exact quantization conditions of the relativistic Toda lattice \eqref{eq:H-N=2} in the simplest case $N=2$ in terms of the refined topological string on the local $\mathbb{P}^1\times\mathbb{P}^1$ geometry in the Nekrasov-Shatashvili limit.
Our focus was on the implication of the S-duality $\hbar\leftrightarrow \wt{\hbar}=4\pi^2/\hbar$ 
when the quantum parameter $q=\re^{\ri\hbar}$ is a root of unity.

We first demonstrated that the $n$-th eigenvalues $E_n$ and $\wt{E}_n$ with the quantum parameter $\hbar=2\pi a/b$ and $\wt{\hbar}=2\pi b/a$, respectively, 
satisfy a simple polynomial relation $P_{a/b}(E_n,R)=P_{b/a}(\wt{E}_n,\wt{R})$, where $P_{a/b}(E,R)$ was defined in \eqref{eq:P-def}.
We then showed that the quantum A-period can be determined exactly in terms of $P_{a/b}(E,R)$, see \eqref{eq:t-exact} and \eqref{eq:t-exact-R=1}.
Interestingly, we found that the polynomial relation above controls Hofstadter's butterfly
and that it has all the information on the spectrum of the Harper equation. 
We also showed that the imaginary part of the derivative of the quantum A-period is exactly the density of states of Hofstadter's Hamiltonian.
In some sense, the correspondence here is natural, since both models have the same underlying symmetry $\cU_q(sl(2, \mathbb{R}))$ 
\cite{Kharchev:2001rs, WZ, NPB_Wiegmann_Zabrodin, Faddeev:1993uk}.
To the authors' knowledge, on the Hofstadter side, the relevance of the modular double property has not been recognized in the literature.

There are many immediate further directions of study.
Firstly, the relation between the exact quantization conditions and the enumerative geometry of the local Calabi--Yau is not just restricted to the case of the local $\mathbb{P}^1\times\mathbb{P}^1$ treated in this paper. We can consider a more general relativistic Toda lattice with more particles \cite{Hatsuda:2015qzx} or a more general completely integrable systems of Goncharov and Kenyon \cite{Goncharov:2011hp} corresponding to general local toric Calabi--Yau manifolds \cite{Franco:2015rnr}.
We should be able to generalize our analysis of the implication of S-duality to these systems. 

Secondly, by multiplying the exponents of the Hamiltonian by the imaginary unit~$\ri$, we have variants of Hofstadter's Hamiltonian for each of the integrable systems just mentioned. We expect that the structure of the spectrum of these generalized versions of Hofstadter's Hamiltonian still controls the analytic structure of the quantum A-period, and that its imaginary part is related to the density of states. We should be able to check these features. 

Thirdly, we can be more ambitious. Note that the determination of the density of states $\rho(E) \rd E$ of Hofstadter's Hamiltonian was quite straightforward, 
once we notice that the density is uniform in the $k$-space: $\rho(E) \rd E \propto \rd k_x \rd k_y$.   
If  the relation between the quantum A-period and the density of states is generic, this observation suggests that the quantum A-period for generic systems, when $q$ is a root of unity, can be readily computed in this manner. 
If the quantum B-period\footnote{Strictly speaking, the quantum B-period is not well-defined for $q$ a root of unity. 
In this case, we need to consider the combination of the B-period and its modular dual.}
can similarly be computed, this would give an independent method to determine the exact quantization condition for the general integrable systems mentioned above, and would also determine the enumerative invariants of the corresponding local Calabi--Yau spaces.

Finally, we should admit that so far the relation to quantum geometry we explored in this paper did not shed any new light on the physics of Hofstadter's system. Rather, we just used the knowledge of Hofstadter's system as an input. 
As the implication of the S-duality $\hbar\leftrightarrow \wt{\hbar}=4\pi^2/\hbar$ on Hofstadter's system does not seem to be extensively studied in the literature, at least to the authors' knowledge, there is a chance that  something new can be said about this issue. 
For example, can we find the unknown function $g(E)$ in \eqref{eq:fractal-generator}, thus explicitly determining the fractal generator?

The authors would hope to come back to some of these issues in the future.

\acknowledgments{
The authors thank Antonio Sciarappa, Futoshi Yagi, and Nobuyuki Yoshioka for useful discussions.
The work of YH is supported in part by the Fonds National Suisse, subsidies 200021-156995 and by the NCCR 51NF40-141869 
``The Mathematics of Physics'' (SwissMAP).
The work of HK is supported in part by JSPS Grant-in-Aid for Scientific Research No. 15K17719 and No. 16H00985. 
The work of YT is partially supported in part by JSPS Grant-in-Aid for Scientific Research No. 25870159,
and  by WPI Initiative, MEXT, Japan at IPMU, the University of Tokyo.
}

\appendix

\section{Period integrals}\label{sec:period}
In this appendix, we briefly review the computations of classical and quantum (A-)periods in the local $\mathbb{P}^1 \times \mathbb{P}^1$.
\subsection{Classical periods}
Let us first consider the classical periods.
It is well-known that special geometry of local Calabi--Yau manifolds
is governed by the Picard--Fuchs (PF) equations.
In the case of local $\mathbb{P}^1 \times \mathbb{P}^1$,
the PF operators are
\be
\ba
\cL_1&= z_2(1-4z_2)\xi_2^2-4z_1^2\xi_1^2-8z_1 z_2 \xi_1 \xi_2 -6z_1 \xi_1+(1-6z_2)\xi_2, \\
\cL_2&= z_1(1-4z_1)\xi_1^2-4z_2^2\xi_2^2-8z_1 z_2 \xi_1 \xi_2 -6z_2 \xi_2+(1-6z_1)\xi_1,
\ea
\ee
where $z_1$ and $z_2$ are complex moduli and $\xi_i=\pd/\pd z_i$.
The classical periods must be annihilated by these operators, i.e.,
solutions to the PF equations.
The important fact is that there are three kinds of singularities in the moduli space:
the large radius point, the conifold point and the orbifold point.
The PF equations allow us to construct the solutions around these singularities (see \cite{Drukker:2010nc}, for instance).
Here, we consider only the large radius point.

The large radius point corresponds to $z_1=z_2=0$.
The solution to the PF equations is constructed by the Frobenius method.
The fundamental period is given by
\be
w_0(z_1,z_2;\rho_1,\rho_2)=\sum_{k,\ell} \frac{\Gamma(2k+2\ell+2\rho_1+2\rho_2)\Gamma(1+\rho_1)^2\Gamma(1+\rho_2)^2}
{\Gamma(2\rho_1+2\rho_2)\Gamma(1+k+\rho_1)^2 \Gamma(1+\ell+\rho_2)^2}z_1^{k+\rho_1}z_2^{\ell+\rho_2}.
\ee
Then the so-called A-periods are given by
\be
-t_i= \frac{\pd}{\pd \rho_i} w_0(z_1,z_2;\rho_1,\rho_2) \biggr|_{\rho_1=\rho_2=0},\qquad
i=1,2.
\ee
It is easy to see that this can be written as
\be
\ba
-t_1=\log z_1 +\wt{\Pi}_A^{(0)}(z_1,z_2),\qquad
-t_2=\log z_2 +\wt{\Pi}_A^{(0)}(z_1,z_2),
\ea
\ee
where
\be
\wt{\Pi}_A^{(0)}(z_1,z_2)=\sum_{(k,\ell) \ne (0,0)} \frac{2\Gamma(2k+2\ell)}{\Gamma(1+k)^2\Gamma(1+\ell)^2}z_1^k z_2^{\ell}.
\label{eq:Pi-A-LR}
\ee
In a similar way, one can construct the B-periods.%
\footnote{Here we refer to the solutions with the logarithmic divergence in $z_i \to 0$ as the A-periods,
while the solutions with the double logarithmic divergence as the B-periods.}  
In our identification \eqref{eq:complex-moduli}, we have
\be
t_2-t_1=\log (z_1/z_2)=-\log R^4.
\ee
Therefore we parametrize $t_1=t$ and $t_2=t - \log R^4$.
Then the parameter $t$ is given by
\be
-t=\Pi_A^{(0)}(E, R)=\log z+\wt{\Pi}_A^{(0)}(z, R^4 z),\qquad z=\frac{1}{E^2}.
\ee
The same result is obtained by the direct period integral of the mirror curve \eqref{eq:mirror-curve-2}.
It turns out that the classical A-period $\Pi_A^{(0)}(E, R)$ is given by
\be
\Pi_A^{(0)}(E, R)=\frac{2}{\pi\ri} \int_{x_-}^{x_+} \rd x \arccosh \( \frac{E}{2}-R^2 \cosh x \),
\label{eq:classical-A-int}
\ee
where $x_{\pm}>0$ are determined by
\be
x_\pm = \arccosh \left[ \frac{1}{R^2} \( \frac{E}{2} \pm 1 \) \right].
\ee
The derivative of $\Pi_A^{(0)}(E, R)$ can be written in closed form.
After a change of variable, one finds
\be
\frac{\pd}{\pd E}\Pi_A^{(0)}(E, R)=-\frac{1}{\pi}
\int_{\frac{E}{2}-1}^{\frac{E}{2}+1}
\frac{\rd t}{\sqrt{(t^2-R^4)(t-\frac{E}{2}+1)(\frac{E}{2}+1-t)}}
\ee
This integral can be performed exactly, and one finally obtains
\be
\frac{\pd t}{\pd E}=\frac{2}{\pi} \( \frac{E^2}{4}-(1-R^2)^2 \)^{-1/2} \eK \( \frac{16R^2}{E^2-4(1-R^2)^2} \),
\label{eq:classical-A}
\ee
where the complete elliptic integral of the first kind is defined by \eqref{eq:ellipticK}.
For $R=1$, one can perform the double sum \eqref{eq:Pi-A-LR} directly:
\be
-t=\log z+4z\,{}_4F_3\(1,1,\frac{3}{2},\frac{3}{2};2,2,2; 16z\).
\ee

\subsection{Quantum periods}
The quantum deformed periods can be computed from the quantized mirror curve:
\be
Q(x+\ri \hbar)+Q(x-\ri \hbar)=(E-2R^2 \cosh x) Q(x).
\label{eq:q-mirror-curve}
\ee
We first consider the semiclassical analysis in $\hbar \to 0$.
In this limit, we take the WKB ansatz
\be
Q(x)=\exp \left[ \frac{\ri}{\hbar} \int^x \rd x' P(x') \right], \qquad
P(x)=\sum_{n=0}^\infty \hbar^{n} P_n(x).
\ee
Plugging this ansatz into the difference equation \eqref{eq:q-mirror-curve},
one can fix $P_n(x)$ order by order.
Note that one obtains two solutions at the leading order $n=0$.
This is because the difference equation has two independent solutions.
For our purpose, either of the solutions is sufficient in order to construct the quantum periods.
It was proposed in \cite{Mironov:2009uv,Mironov:2009dv} that the quantum A-period are obtained by
\be
\Pi_A(E, R; \hbar)=\sum_{n=0}^\infty \hbar^{2n} \Pi_A^{(n)}(E, R),\qquad
\Pi_A^{(n)}(E,R)= \oint_A \rd x \, P_{2n} (x).
\ee
where the integral contour $A$ should be chosen as a closed circle around two points $x_\pm$.
Of course, at the leading order, the integral reduces to the classical one \eqref{eq:classical-A-int}
(up to an irrelevant rescaling).
In this way, one can compute analytic forms of the quantum corrections order by order,
but this method only gives the period perturbatively in $\hbar$.
The quantum B-period can be computed by changing the integration contour appropriately.

Another powerful method was proposed in \cite{Aganagic:2011mi}.
We first rewrite the difference equation \eqref{eq:q-mirror-curve} as
\be
V(X)+\frac{1}{V(q^{-1}X)}=E-R^2\(X+\frac{1}{X}\),\qquad X=\re^x,\quad q=\re^{\ri \hbar},
\label{eq:q-diff}
\ee
where $V(X)=Q(q X)/Q(X)$.
We solve this equation in the large $E$ limit.
The right hand side behaves as $E$ in $E \to \infty$.
There are two possibilities: $V(X) \sim E$ or $V(q^{-1}X) \sim E$.
If $V(X) \sim E$, then $1/V(q^{-1}X) \sim E^{-1}$, and vice versa.
It is sufficient to consider the first case.
We can take the ansatz
\be
V(X)=E-R^2\(X+\frac{1}{X}\)-\sum_{k=1}^\infty \frac{v_k(X;q)}{E^{k}}.
\ee
The coefficients $v_k(X;q)$ can be easily fixed by the $q$-difference equation \eqref{eq:q-diff}.
The first few results are
\be
\ba
v_1(X;q)&=1,\qquad v_2(X;q)=R^2(q^{-1}X+qX^{-1}),\\
v_3(X;q)&=R^4(q^{-1}X+qX^{-1})^2+1.
\ea
\ee
Thus the expansion of the logarithm is
\be
\log V(X)=\log E -\frac{R^2(1+X^2)}{XE}-\frac{2X^2+R^4(1+X^2)^2}{2X^2 E^2}+\cO(1/E^3).
\ee
The claim in \cite{Aganagic:2011mi} is that the quantum A-period in $E \to \infty$ is given by
\be
\ba
\wt{\Pi}_A(E, R; q)&=-\Res_{X=0} \frac{2}{X}\log \frac{V(X)}{E} \\
&=\Res_{X=0} \( \frac{2R^2(1+X^2)}{X^2E}+\frac{2X^2+R^4(1+X^2)^2}{X^3 E^2}+\cO(1/E^3) \)
\ea
\ee
In this way, one obtains the expansion \eqref{eq:quantum-A}.
In a similar manner, one can also compute the quantum B-periods,
but the computation is much more complicated.
See \cite{Aganagic:2011mi} in detail.

An open problem in these computations is the following.
In the semiclassical computation, one obtains the quantum periods around $\hbar=0$.
Since each coefficient is exact in $z$ (or $E$), one can analytically continue it to
the whole $z$-plane. However the result is perturbative in $\hbar$, and we have to
resum it if we want to know the behavior for finite $\hbar$.
In general, the semiclassical expansion is asymptotic, and we have a delicate resummation
problem (Borel summability, ambiguity of resummation, etc.).

On the other hand, the method in \cite{Aganagic:2011mi} gives the quantum periods around the large radius point $z=0$ but exact in $\hbar$.
This expansion is convergent for $|q|=1$.
If we want to analytically continue the periods outside the convergence regime,
we again encounter another resummation problem.
Though the expansion around $z=0$ is convergent,
its resummation seems technically very difficult.
In other words, we do not know any systematic ways to compute the quantum periods far from the large radius point, at least so far.

We would like to emphasize, in the main text of this paper, we partially solved this problem.
We found the explicit analytic expression of the quantum A-period for any $q$ of the form $\re^{2\pi \ri a/b}$,
and its analytic structure turned out to be quite complicated.

\bibliographystyle{ytphys}
\bibliography{Hofstadter-v2}

\end{document}